\documentclass[12pt]{iopart}

 \usepackage{graphicx} 
 \usepackage{cite}
\usepackage[colorlinks, linkcolor = blue, citecolor = blue, urlcolor = blue]{hyperref}

\begin{document}

\title[\footnotesize Quantum Imaging of Photonic Spin Texture in an OAM Beam with NV Centers in Diamond] {Quantum Imaging of Photonic Spin Texture in an OAM Beam with NV Centers in Diamond}

\author{Shoaib Mahmud$^{1,2}$, Wei Zhang$^{1,2}$, Farid Kalhor$^{1}$, Pronoy Das$^{1,2}$, and Zubin Jacob$^{1,2}$}

\address{$^{1}$Elmore Family School of Electrical and Computer Engineering, Purdue University, West Lafayette, Indiana 47907, USA}

\address{$^{2}$Birck Nanotechnology Center, and Purdue Quantum Science and Engineering Institute, West Lafayette, Indiana 47907, USA}

\ead{zjacob@purdue.edu}
\vspace{10pt}
\begin{indented}
\item[]February 2025
\end{indented}

\begin{abstract}
Photonic spin texture (PST), the spatial distribution of the spin angular momentum (SAM) of light, is connected to unique properties of light, such as optical skyrmions and topological optical N-invariants. There has been recent progress on the generation and manipulation of PST using various methodologies. However, a challenge remains for the sub-wavelength characterization of PST. Here, we demonstrate nitrogen-vacancy (NV) centers in diamond as nanoscale quantum sensors for imaging the PST of a beam with orbital angular momentum (OAM). Leveraging the coherent interaction between photon spin and NV center electron spin at cryogenic temperature (77 K), and using the Hahn-Echo magnetometry technique, we experimentally demonstrate the imprinting of the PST on the quantum phase of NV centers. Our work can lead to the development of a quantum imaging platform capable of characterization of the spin texture of light at sub-wavelength scales.
\end{abstract}

%
%
%
%
%

\section{Introduction}

Structured light with a unique PST profile holds potential applications across fields of photonics \cite{wang2023photonic,shen2024optical,lei2025skyrmionic}. Optical beams with controlled spin textures are essential for the study of the topological properties of light, which has applications in noise-resilient optical communication, superresolution imaging, and quantum metrology \cite{du2019deep,chen2021engineering,yang2021non,rubinsztein2016roadmap, wang2022topological}. Moreover, exotic phases of light characterized by spin-orbit coupling \cite{bliokh2015spin}, quantum spin hall effect\cite{bliokh2015quantum}, spin-momentum locking \cite{van2016universal} and topological electromagnetic field distribution \cite{van2018photonic, van2021optical} are fundamentally linked to the spin texture of light. While SAM has been extensively studied over the years, its distribution in amplitude and polarization (PST) has recently drawn attention. This distribution can be precisely controlled by modulating the phase of the electromagnetic field in the optical beam \cite{shen2022generation, gao2020paraxial}. A gradual phase change in the azimuthal direction induces orbital angular momentum (OAM) in the laser beam \cite{padgett2004light, shen2019optical,das2024quantum2,bozinovic2013terabit}. The modulation of spin texture through OAM reveals various topological applications of light \cite{shen2024optical, sugic2021particle, he2022towards}. Furthermore, spin texture can be altered by interconversion between OAM and SAM via spin-orbit coupling \cite{du2019deep, dai2022ultrafast}. Given the extensive research on manipulating spin textures of optical beams, there is a need for a nanoscale, non-invasive, on-chip integrable quantum sensor capable of sub-wavelength characterization of spin texture in structured light.  

Unlike optical intensities, the polarization singularities and local spin states of light are not limited by diffraction. As a result,  PST requires a nanoscale sensing tool to achieve accurate sub-wavelength characterization and imaging. Some state-of-the-art methods for characterizing PST include near-field scanning optical microscopy \cite{rotenberg2014mapping, yin2020spin}, pump-probe methodology \cite{davis2020ultrafast}, nanoparticle scattering \cite{neugebauer2018magnetic}, or surface relief structuring \cite{tamura2024direct}. NV center in diamond has recently shown its capability in revealing the topological magnetic spin textures in antiferromagnet utilizing quantum magnetometry \cite{tan2024revealing}. NV centers have also demonstrated high resolution and sensitivity as quantum imaging platforms for magnetic skyrmions \cite{dovzhenko2018magnetostatic}, magnetic fields \cite{ku2020imaging, cujia2022parallel}, crystal stress \cite{kehayias2019imaging}, living biological samples \cite{le2013optical} etc. Furthermore, these nanoscale quantum sensors can work with high sensitivity at a broad range of temperatures and pressures. The sensing capability of such a quantum sensor has not yet been demonstrated for imaging the spin texture in structured light. 

\begin{figure*}[!t]
\centering\includegraphics[scale = 0.5]{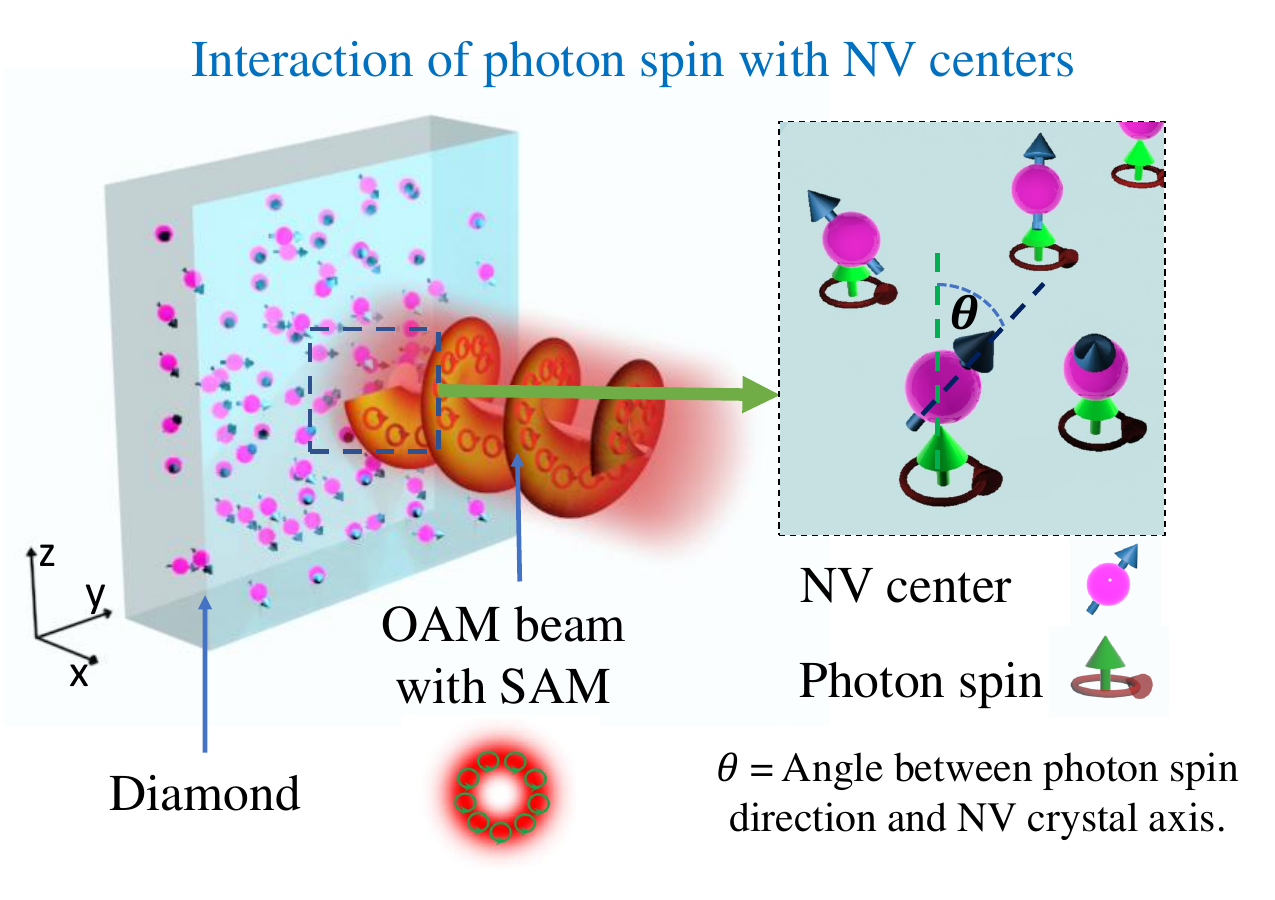}
\caption{ The interaction of the PST in an OAM beam with NV centers. The PST is controlled by the global helical phase-front of the OAM beam and donut-shaped beam profile due to the phase singularity at the center. The local photon spin of the beam interacts with NV centers in the diamond sample.}
\label{Figure1}
\end{figure*}

In this work, we present NV centers in diamond as a nanoscale quantum imaging tool for the spin texture of light. The recent demonstration of the coherent interaction between photonic spin and NV centers suggests a pathway for imaging the spin texture of optical beams \cite{kalhor2021quantum, kalhor2022optically}. Here, we leverage this coherent interaction to achieve sub-wavelength characterization of the spin texture of OAM beams. When a PST beam is irradiated on the diamond, the NV centers interact with the effective static magnetic field generated by the local polarized light fields (Fig. \ref{Figure1}). This interaction leads to a phase change of the NV centers which is then readout optically. We experimentally reveal the spin texture in an OAM beam, measuring the sign and magnitude of the PST-induced phase shift of the NV centers using a Hahn-Echo magnetometry technique. 

\section{PST imprinted phase readout on NV centers spin qubits}

The definition of PST can be understood from the total angular momentum equation of an electromagnetic field. In the presence of both SAM and OAM, the total angular momentum of an electromagnetic field is given by, $\vec{J}=\epsilon\int d^3 r \ \space\vec{\cal E}\times\vec{A}+\epsilon\int d^3 r \ {\cal E}^j(\textbf{r}\times\nabla) A^j$, where  $\vec{\cal E}(\vec{r},t)$ and $\vec{A}(\vec{r},t)$ are the electric field and the vector potential respectively, and $\epsilon$ is the permittivity \cite{das2024quantum, yang2022quantum}. The first term of the equation corresponds to the SAM and the second term corresponds to the OAM of light. The PST is associated with the integral kernel of the SAM term. This spin texture, represented by the term, $\vec{S}_E = \epsilon \ \vec{\cal E}(\vec{r},t)\times \vec{A}(\vec{r},t)$ is imprinted on the quantum phase of NV centers in our imaging method. OAM being the global rotation along the beam axis, can change the field intensity and orientation profile of the spin texture.

\begin{figure}[h!]
\centering\includegraphics[scale = 0.4]{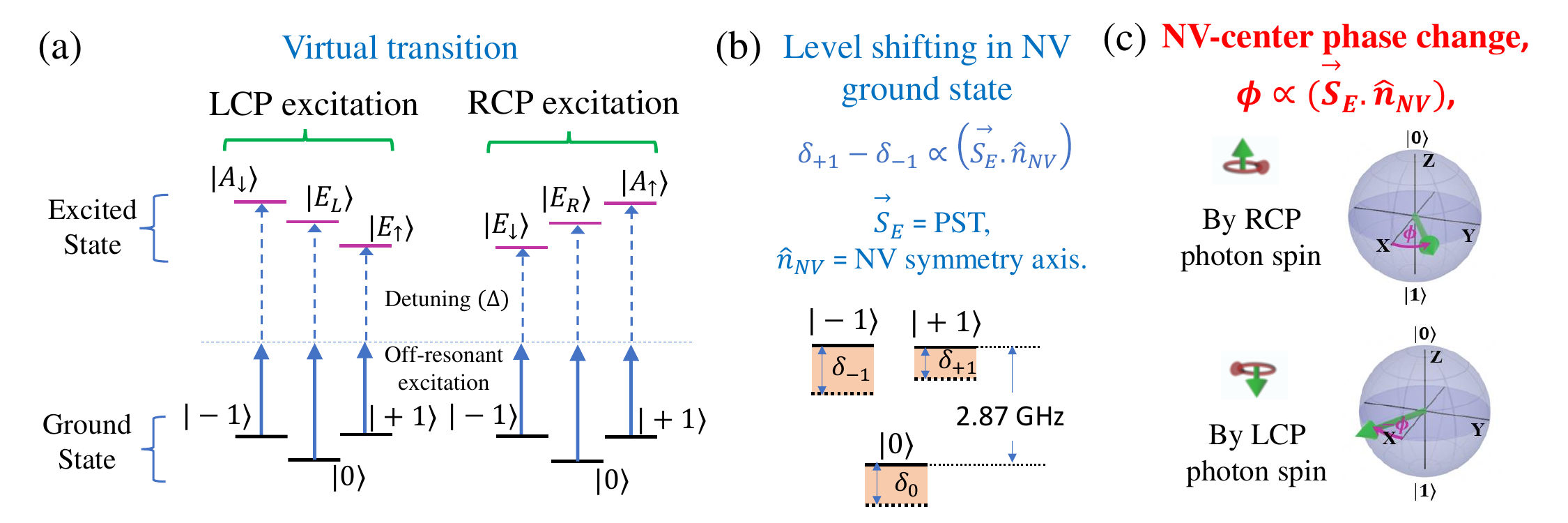}
\caption{PST induced phase change on NV centers. (a) The effect of an off-resonant excitation by a PST beam on the ground state of NV centers. The off-resonant excitation induces a virtual electric dipole transition between the ground states and excited states. (b) This virtual transition causes a spin-selective ac stark shift induced level shifting in the NV center ground state. The strength of this level shifting is proportional to the projection of PST along NV symmetry axis. (c) The virtual level shifting causes a phase change in the superposition state of NV centers. By measuring the phase change optically, we can quantify the effect of PST.}
\label{Figure2}
\end{figure}

Our quantum sensing approach relies on a spin-selective level shifting in the ground state of NV centers through the interaction with photon spin \cite{kalhor2022optically}. An off-resonant photon spin $(\lambda \  \approx \ 660\ nm)$ interacts with NV center electron spin, causing a virtual shifting in the ground state energy levels of NV centers (Fig. \ref{Figure2}). The strength of this interaction depends on the spin texture of the beam. The ground state and excited states energy levels of NV centers can be written in terms of the orbital and spin-resolved states as shown in Fig. \ref{Figure2}(a) \cite{kalhor2021quantum,maze2011properties}. The ground state consists of the energy levels $\vert0\rangle$, $\vert-1\rangle$ and $\vert+1\rangle$, while the excited energy eigen-states include $\vert E_L\rangle=\frac{i}{2}[\vert a_1\rangle\vert e_+\rangle-\vert e_+\rangle\vert a_1\rangle]\bigotimes\vert0\rangle$, $\vert E_R\rangle=\frac{i}{2}[\vert a_1\rangle\vert e_-\rangle-\vert e_-\rangle\vert a_1\rangle]\bigotimes\vert0\rangle$, $\vert E_\downarrow\rangle=\frac{i}{2}[\vert a_1\rangle\vert e_-\rangle-\vert e_-\rangle\vert a_1\rangle]\bigotimes\vert-1\rangle$, $\vert E_\uparrow\rangle=\frac{i}{2}[\vert a_1\rangle\vert e_+\rangle-\vert e_+\rangle\vert a_1\rangle]\bigotimes\vert+1\rangle$, $\vert A_\uparrow\rangle=\frac{i}{2}[\vert a_1\rangle\vert e_-\rangle-\vert e_-\rangle\vert a_1\rangle]\bigotimes\vert+1\rangle$ and $\vert A_\downarrow\rangle=\frac{i}{2}[\vert a_1\rangle\vert e_+\rangle-\vert e_+\rangle\vert a_1\rangle]\bigotimes\vert-1\rangle$. Here, $\vert e_\pm\rangle=\mp(\vert e_x\rangle \pm i\vert e_y\rangle)/\sqrt{2}$ and $\vert a_1\rangle$, $\vert e_x \rangle$, $\vert e_y\rangle$ are the orbital states and $\vert0\rangle$, $\vert-1\rangle$, $\vert+1\rangle$ are the spin states of NV center energy levels. The off-resonant excitation produced by our PST beam causes the spin-selective virtual level transitions between the ground and excited states as depicted in Fig. \ref{Figure2}(a). 

These virtual level transitions cause the ground state energy level to shift in the form of AC stark shift. The strength of these shifts can be expressed by the equation -
\begin{equation}
    \delta_i =\frac{1}{4\hbar^2}\sum_f\frac{\vert\langle f \vert \vec{d} \vert i \rangle.\vec{E}(\vec{r},t)\vert^2}{\Delta_{if}+\Gamma_f^2/4\Delta_{if}}
    \label{equation1}
\end{equation}

Here, $\vert i\rangle$ and $\vert f\rangle$ are the initial and final states of the possible transitions from ground state to the excited state of NV centers, $\Delta_{if}$ is the detuning between the transition resonance frequency and the off-resonant excitation frequency, $\vec{E}(\vec{r},t)$ represents the electric field and $\Gamma_f$ is the spontaneous decay rate of the final state. The transition element, $\langle f \vert \vec{d} \vert i \rangle$ indicates the strength of the optical transition from the ground state to the excited state. The strength of this optical transition depends on the PST of the light and the spin states of the NV centers. This gives rise to the PST-dependent phase shift, $\phi$ in the superposition state of NV centers which can be written as (Fig. \ref{Figure2}(b)) - 
\begin{equation}
\phi\propto(\delta_{+1}-\delta_{-1})\propto(\vec{S}_E.\hat{n}_{NV})
\label{equation3}
\end{equation}

Equation \ref{equation3} shows that the relative level shifting between $\vert+1\rangle$ and $\vert-1\rangle$ states has a dependence on $\vec{S}_E$ and the direction of NV crystal symmetry axis, $\hat{n}_{NV}$. The relative level shifting causes a phase change in the superposition of NV centers (Fig. \ref{Figure2}(c)) \cite{buckley2010spin}, which can be measured optically to quantify the PST in the OAM beam. Through this coherent interaction, the spin texture of light is imprinted on the phase of NV centers. We exploit this coherent light-matter interaction to image the spin texture of the OAM beam. 

\begin{figure*}[h!]
\centering\includegraphics[scale = 0.4]{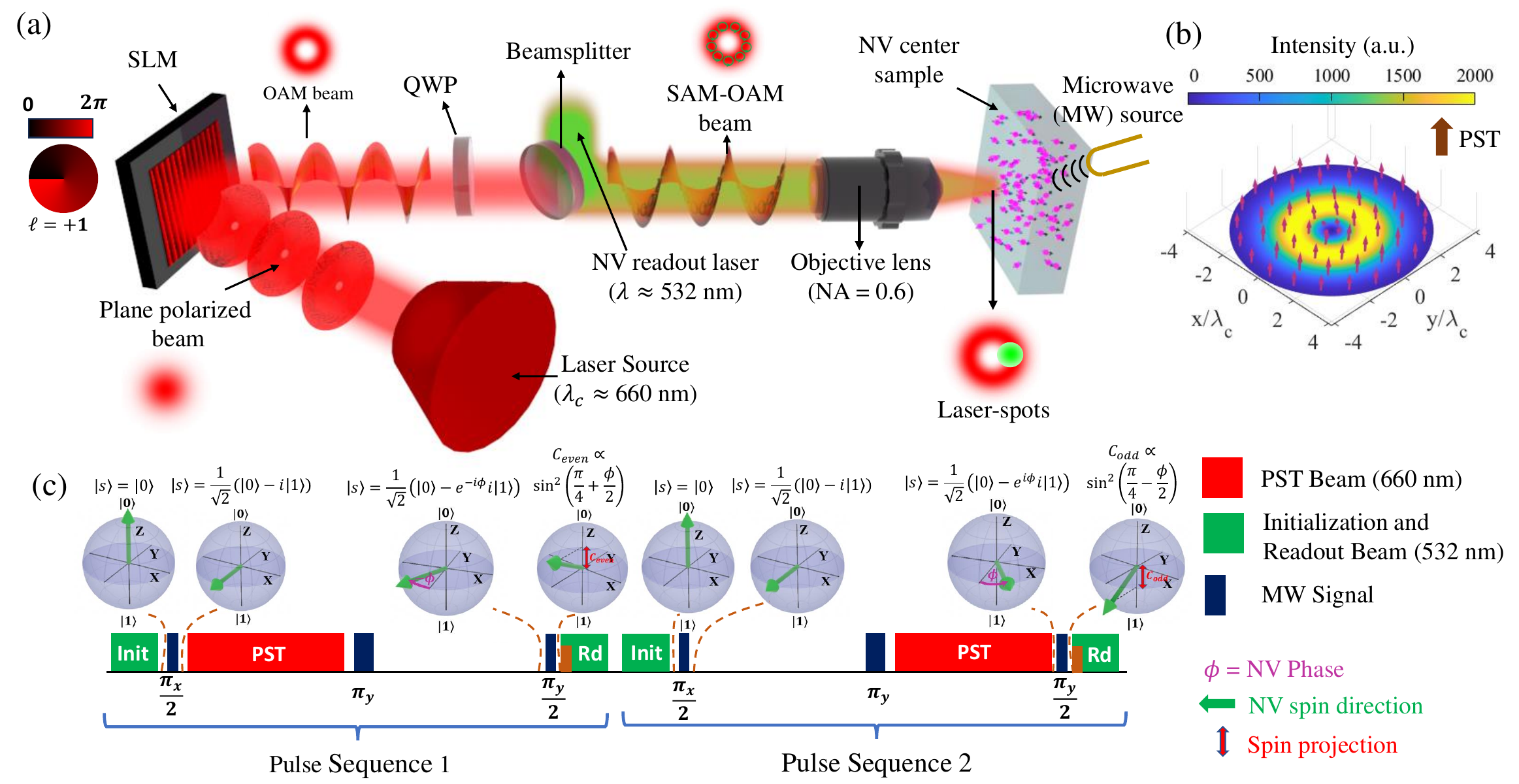}
\caption{Experimental setup and measurement protocol for imaging the PST in an OAM beam. (a) The experimental setup for imaging the spin texture in the OAM beam. The SLM-generated spatial phase gradient in the incident plane-polarized beam induces OAM. A green laser with around 500 nm spot size is used for reading out the phase of NV centers at different positions of the incident OAM beam. A MW signal is used for coherent driving of the NV center spin. (b) The simulated spin texture of OAM beam with topological charge $\ell=1$. (c) The pulse sequence that has been used to measure the phase change in the NV centers. The position of the NV center spin on the Bloch sphere at different stages of the pulse sequence is depicted. Initially, the NV center is brought into a superposition state using a MW pulse, followed by the application of a phase-shifting (PST) beam to induce a net phase change in the superposition state. The $\pi$-pulse within the Hahn echo pulse sequence is applied to cancel any external unmodulated signals during our measurements. After phase accumulation, the readout is performed by projecting the superposition state to a population difference.}
\label{Figure3}
\end{figure*}

For the experimental realization of the quantum imaging of the PST in an OAM beam, we use the optical setup as shown in Fig. \ref{Figure3}(a) (more details in \ref{Experimental setup}). A horizontally linear-polarized laser output from an external cavity diode laser (ECDL) system is passed through a spatial light modulator (SLM) ($1920\times1200$ XY Phase Series, pixel size of $8.0 \times 8.0 \ \mu m^2$, Meadowlark Optics). The SLM can produce a spatially varying phase profile in the electric field of the beam controlled by the computer-generated hologram. This spatial variation in phase profile induces the OAM with the designed parameter within the beam. Subsequently, the linearly polarized output from the SLM passes through a quarter-wave plate (QWP) at an angle of $\theta$, allowing control over the spin texture within the OAM beam. This free space OAM-SAM optical beam is then incident on a diamond sample with bulk NV concentration of around $300\ ppb$ through an objective lens of 0.6 numerical aperture (NA). 

We conducted beam characterization and spin texture simulation to achieve accurate PST imaging and to carefully analyze the NV-center phase change when irradiated by the OAM beam.  In Fig. \ref{Figure3}(b), we have plotted the simulated intensity pattern and spin texture of the OAM beam incident on NV centers. The dimension of the beam is obtained from the experimentally generated confocal images using our microscopy setup as shown in Fig. \ref{Figure4}. These confocal images are based on the photon counts obtained from the reflection of the diamond surface. From these images, we can extract the beam properties (waist diameter, electric field pattern, etc.). We use these experimental parameters to generate the simulated intensity and spin texture profile of the OAM beam (details in \ref{OAM beam generation}). 

Following the beam characterization, the Hahn-echo AC magnetometry technique is used to measure the PST-induced phase change in NV centers (details on setup of the measurement protocols and parameters in \ref{Setup of the measurement protocols}). Fig. \ref{Figure3}(c) shows the pulse sequence of our magnetometry protocol and the NV center spin states at different positions of the sequence. The initialization and readout of the NV center spin states are done using a  $\lambda \approx 532\ nm$ wavelength laser (spot-size $\approx 500\ nm$  and power $\approx$ 100 $\mu W$). After initializing the NV center spin to $\vert0\rangle$ state, we use the first MW $\frac{\pi}{2}-$pulse to bring the spin to a superposition state, $\frac{1}{\sqrt{2}}(\vert0\rangle-i\vert1\rangle)$. The subsequent MW $\pi-$pulse (length $\approx 30\ ns$) is to decouple any nearby noise sources from our measurement. The PST beam is applied in the odd and even free precession time slots of two consecutive measurements (length $\approx 20\ \mu s$ and power $\approx 4.5\ mW$). It produces a net phase shift $\phi$ in the NV spin superposition state, $\frac{1}{\sqrt{2}}(\vert0\rangle-e^{-i\phi}i\vert1\rangle)$. The sign and magnitude of this phase shift depend on the field amplitude and polarization of the PST. This phase is projected onto a spin population difference $(C_{even}$ and $C_{odd})$ using a final $\frac{\pi}{2}$-pulse. The population is read out optically using a green pulse. This measurement is repeated by applying the PST beam in the odd precession time slot of the Hahn echo pulse sequence. The subtraction of measurement 1 from 2, mitigates any common-mode sources of noise. Using this Hahn-echo magnetometry technique, we can reach a sensitivity of around $40\ Deg/\sqrt{Hz}$. We scan our readout green laser at different positions of the OAM beam using a set of Galvo mirrors and measure the phase change of NV centers at the corresponding positions. In this way, we image the spin texture of the OAM by reading out the phase information of the NV center spin qubits. All these measurements are conducted at $77\ K$ temperature in order to avoid phonon-assisted off-resonant absorption in the NV centers by the incident PST laser \cite{ulbricht2018vibrational}.

\section{Imaging of the spin texture in an OAM beam}

\begin{figure*}[!t]
\centering\includegraphics[scale = 0.55]{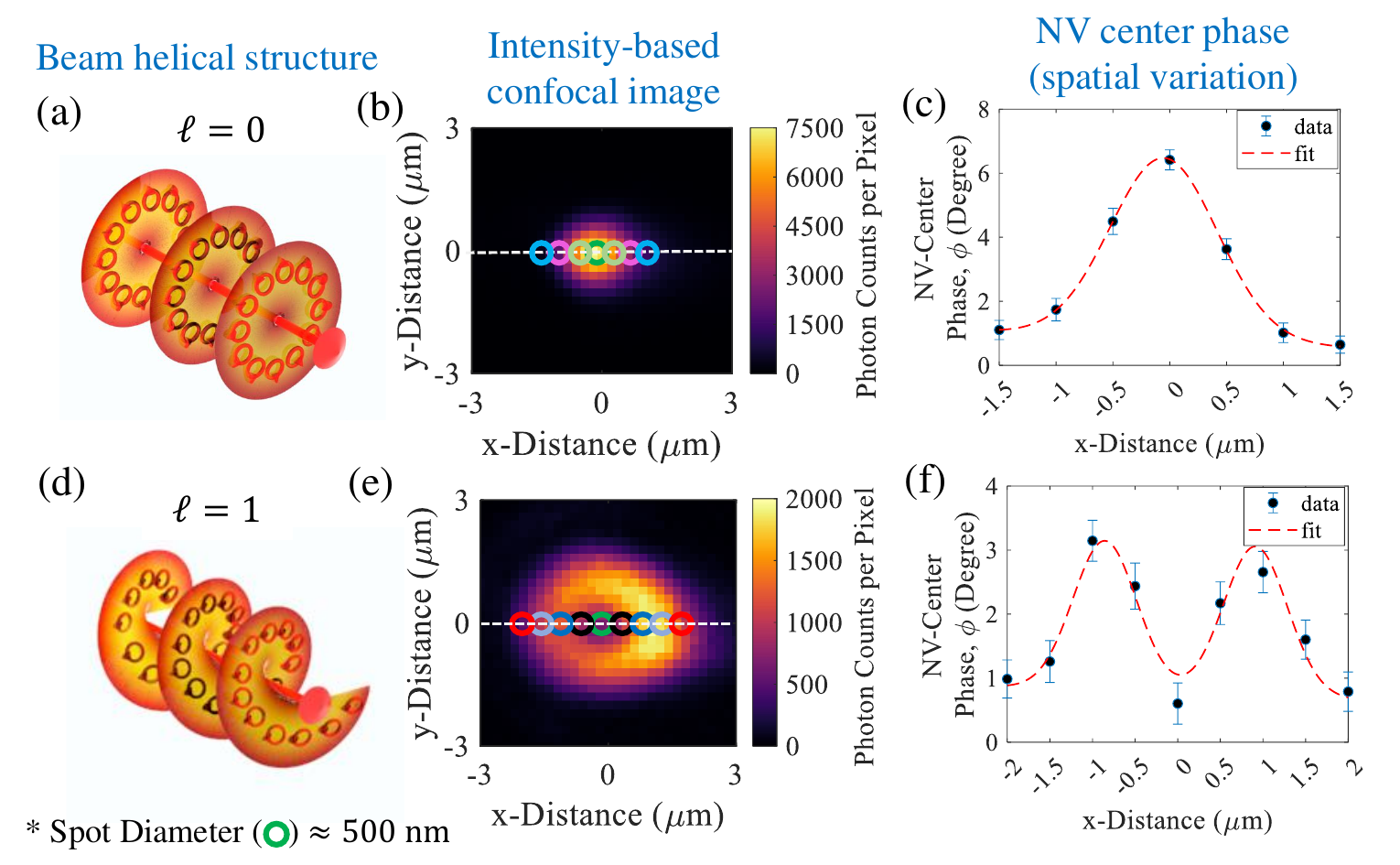}
\caption{Imaging the PST in an OAM beam with varying positions. (a) The Gaussian mode with topological charge $\ell = 0$ consists of propagating plane wavefront. (b) The confocal image shows the beam profile of a Gaussian mode. The colored circle shows the imaging spots for the phase of NV centers. (c) The phase changes of the NV centers at the imaging spots of the Gaussian mode. The phase change is following the intensity pattern of the beam. (d) The helical phasefront arises in the OAM beam with topological charge $\ell = 1$. (e) The confocal image corresponding to OAM mode $\ell=1$ shows the position of measurements of the spin texture of the optical beam. (f) Graph showing the phase change of NV centers due to the interaction with the spin texture of the optical beam with OAM mode $\ell=1$.}
\label{Figure4}
\end{figure*}

\begin{figure*}[!t]
\centering\includegraphics[scale = 0.55]{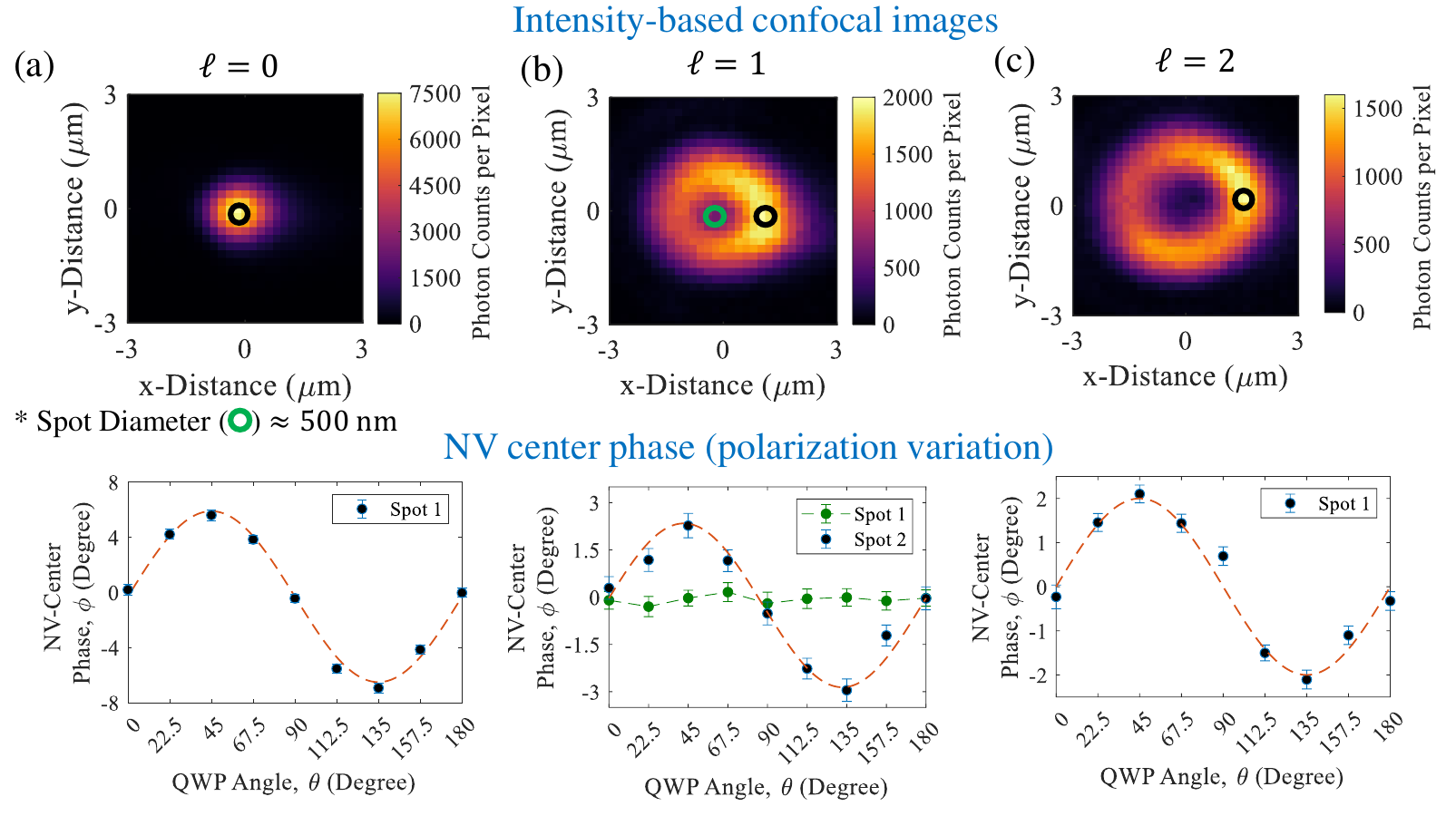}
\caption{Imaging the PST in an OAM beam with varying polarization. (a) The graph depicts the change of the phase of NV centers with the change of the ellipticity of polarization at the center of a Gaussian mode beam. The NV center phase change follows the sinusoidal variation of the PST with the change of the QWP angle. (b) This graph shows the change of the NV center phase with variation of the ellipticity of polarization at phase singularity (spot 1) and max intensity (spot 2) spots of the OAM beam with topological charge $\ell = 1$. (c) The graph illustrates the change of the NV center phase with the change of the polarization of the incident beam at the max intensity point of the OAM beam with topological charge $\ell = 2$. }
\label{Figure5}
\end{figure*}

In order to obtain a complete picture of the PST of a free-space OAM beam, we have done multiple imaging varying the position and polarization of the beam. In monochromatic limit, the PST can be written as, $\vec{S}_E = \frac{\epsilon}{4\omega}\  Im(\vec{E}^*\times\vec{E}) \propto |\vec{E}|^2\ sin(2\theta)\ \hat{n}$ \cite{bliokh2015transverse}, where $\vec{E}$ is the electric field, $\vec{E} = cos\theta\ \hat{x}+i\ sin\theta\ \hat{y}$. According to this equation, the PST depends on the field strength $|\vec{E}|^2$, the ellipticity of polarization $sin(2\theta)$, and the direction of the photon spin, $\hat{n}$. We have shown that using our imaging methodology, we can quantify all the properties of PST by measuring the NV center phase, $\phi\propto \vec{S}_E.\hat{n}_{NV}$.

Fig. \ref{Figure4} shows the imaging of the PST at different positions of the beam along the horizontal direction. Here, the polarization of the beam is set to right-handed circular. The spot-size of the measurement point is around $500\ nm$ (diffraction limitation of readout beam). From our confocal images, it is visible that the OAM beam size varies from $1.4\ \mu m$ to $3\ \mu m$ depending on the OAM number. The smaller wavelength of the readout beam $(\approx 532\ nm)$ compared to the PST beam $(\approx 660\ nm)$ enables us to do the sub-wavelength characterization. This difference in this wavelength scale can be made larger by increasing the wavelength of the PST beam. However, an increase in the wavelength from the resonant wavelength of NV centers $(\approx 637\ nm)$ causes a reduction in the interaction strength between NV centers and PST beam. So, the wavelength of our PST beam is kept at 660 nm for an enhanced interaction strength. For the spatial variation along the x-direction of the Gaussian mode (Fig. \ref{Figure4}(b)), the NV center phase is changing in a Gaussian pattern (Fig. \ref{Figure4}(c)). This is because the phase change of the NV center follows the $|\vec{E}|^2$ component, where the polarization and direction are kept constant. But, in the case of OAM mode $\ell=1$, the presence of phase singularity near the beam axis, results in a donut-shaped intensity profile (Fig. \ref{Figure4}(d,e)). This gives rise to the two side lobes in the NV center phase measurements along the x-direction (Fig. \ref{Figure4}(f)). These measurements show that our imaging methodology is capable of capturing the field distribution $|\vec{E}|^2$ component of the spin texture.

Next, we perform imaging keeping the field strength $|\vec{E}|^2$ fixed, but varying the ellipticity of the polarization $sin(2\theta)$.  Fig. \ref{Figure5}(a,b,c) show the imaging of the PST of the beam with varying polarization for OAM numbers $\ell=0,1,2$ respectively. In these measurements, the field strength ($|\vec{E}|^2$ term) is constant due to the fixed position of the measurement spot. However, the ellipticity of the polarization ($sin(2\theta)$) is varied by tuning the angle of the easy axis of the QWP, $\theta$ with respect to the polarization axis of the incident beam. It is visible from the measurements in Fig. \ref{Figure5} that the NV center phase is following a $sin(2\theta)$ variation of the QWP angle at different spots. This is because the polarization component of the PST is being imprinted on the phase of the NV centers.  These measurements demonstrate that our imaging methodology with NV centers is capable of capturing not only the intensity profile but also the polarization component of the PST. Note that the photon spin direction in the PST beam is also extractable if measured with a single NV. Here, we conducted the PST imaging using an NV ensemble which loses the spin direction information. The resolution of the measurement shown in this work is $500\ nm$ limited by the spot size of the green readout beam. However, our methodology can be projected to capture the vectorial nature of the photon spin with sub-diffraction resolution through vector magnetometry using a single NV. In this way, NV magnetometry can achieve ultra-subwavelength resolution in capturing all the components of the PST using the NV center spin qubits.

\section{Conclusion}
Our measurements show good agreement with the intuition about the spin texture of OAM obtained from the theoretical derivations and simulation results. This supports the role of the NV center as a quantum imaging tool for the spin texture of OAM beams. OAM serves as a basis for generating different complex spin texture profiles in topological quasiparticles like skyrmions, bimeron, hopfion, etc \cite{shen2024optical,sugic2021particle,lei2021photonic}. Our imaging methodology with OAM beams can be used for complete sub-wavelength characterization of the spin features of these topological quasiparticles of light. Our methodology also holds promise for various applications in studying diverse light-matter interactions featuring exotic PST profiles. \\

\section{Acknowledgments}
This work is supported by the Army Research Office (W911NF-21-1-0287) and the U.S. Department of Energy (DE-SC0017717).\\

\section{Data Availability}
All data that support the findings of this study are included within the article (and any supplementary files).\\

\appendix

\section{Experimental setup} \label{Experimental setup}
\begin{figure*}[!h]
\centering\includegraphics[scale = 0.42]{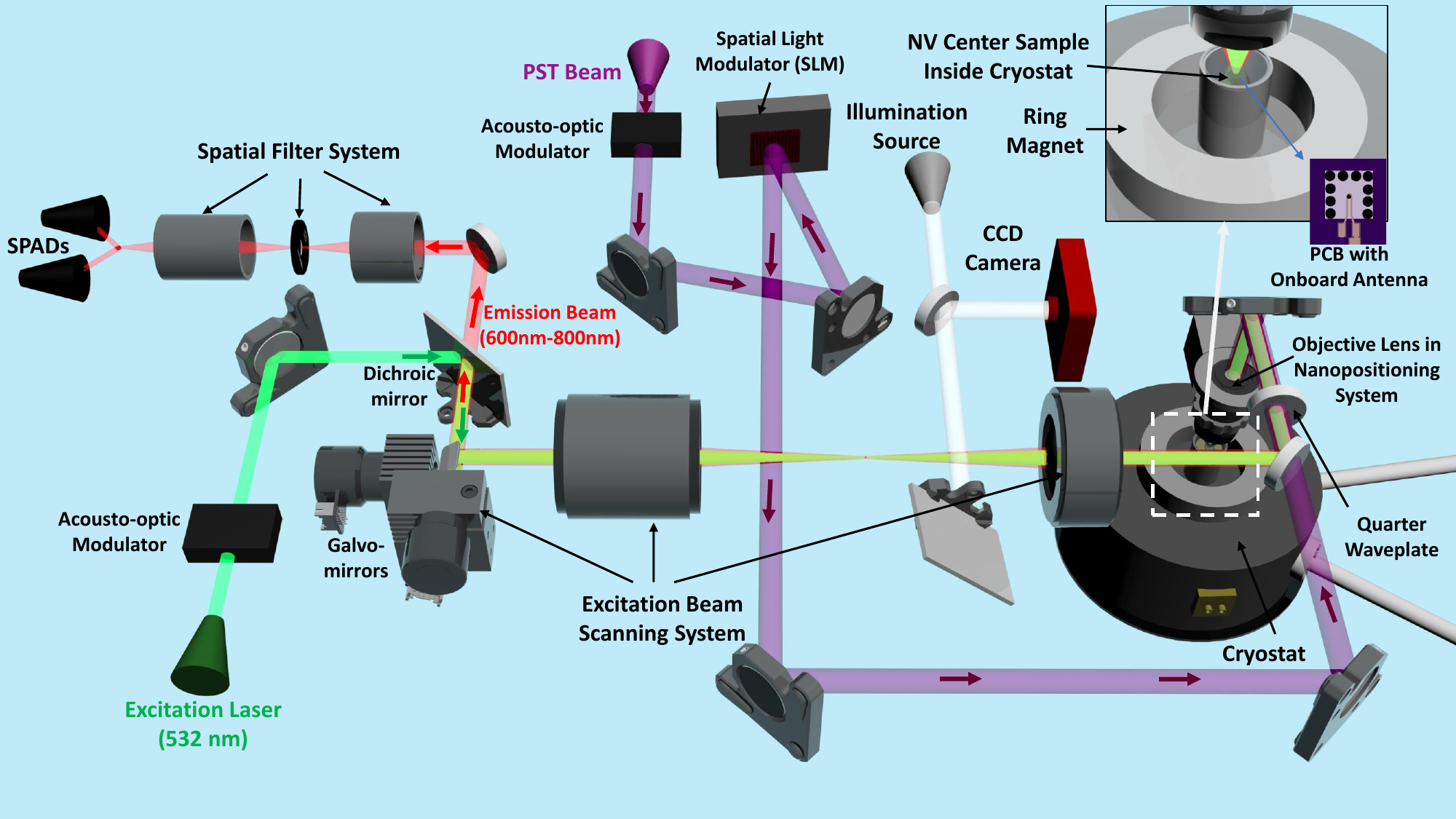}
\caption{The experimental setup designed for quantum imaging of the PST in an OAM beam. It is a home-built confocal setup equipped with imaging capability along all the directions of the diamond sample.}
\label{Figure_S1}
\end{figure*}

We have a home-built confocal microscopy setup, as shown in Fig. \ref{Figure_S1}. It consists of some optical elements arranged in several beam paths and includes a Janis ST-500 microscopy cryostat. The excitation source for the initialization and readout of NV centers is a 532 nm beam generated by a continuous-wave (CW) solid-state laser. For pulsed measurement, this beam is modulated by an acousto-optic modulator (AOM) controlled via a programmable TTL pulse generator. The laser is focused onto the NV center sample using an objective lens with an NA of 0.6. The objective lens has a maximum working distance of 4 mm and is collar-corrected for the 1 mm thick glass window of the cryostat. The lens focuses the green laser to a spot size of approximately 500 nm on the diamond sample. The system for scanning the excitation beam consists of a dual-axis galvo-mirror set, a scanning lens, and a tube lens. These components work together to direct the beam at various angles through the aperture of the objective lens, scanning the laser beam across the XY plane of the sample. A piezo Z-nanopositioner, which holds the objective lens, enables focusing of the beam at different planes of the NV centers sample.

The diamond sample is mounted on the cold finger of the cryostat using a printed circuit board (PCB). A copper antenna on the PCB supplies an MW signal to the NV center sample. Apiezon N thermal grease ensures good thermal contact between the diamond sample and the cold finger. Temperature monitoring of the cold finger is performed using a Lakeshore Cryotronics model-335 temperature controller. Variable magnetic fields are applied to the NV center sample by placing appropriate ring magnets around the cryostat snout. The cryostat operates at 77 K with liquid nitrogen to minimize phonon-assisted off-resonant excitation by the PST beam.  

The PST beam originates from a 660 nm laser source in a Thorlabs laser diode system. The normal output of a laser diode is not narrowband and frequency stable. To achieve a narrowband and frequency-stable output, the laser diode output is passed through an ECDL system. Subsequently, it is passed through an SLM to generate the OAM. The PST beam is aligned with the green excitation laser using a beamsplitter. It is then passed to the diamond sample through a QWP, to tune the SAM of the PST beam. An objective lens focuses the laser to a spot size of around $1.4-3\ \mu m$ (depending on the OAM number) on the diamond sample. The position of the green laser can be scanned on the PST beam to optically read out the NV centers phase.

Photoluminescence (PL) signals emitted from the diamond sample are collected with the same objective lens and pass through a dichroic mirror with a cutoff wavelength of 550 nm. The PL signal is detected using a pair of Micro Photon Devices SPADs. Higher spatial resolution is obtained using a pair of lenses with a pinhole aperture to collect only fluorescence from the objective's focal plane. The time-tagged PL signal is processed by a Hydraharp time-correlated single photon counting (TCSPC) device. This time-tagged signal is used to read the spin states of NV centers during various measurements and to create confocal images.  

Additionally, the setup includes a Thorlabs scientific camera for obtaining brightfield microscopy images from the focal plane of the objective lens. A white light source is used for illuminating the sample surface during the initial alignment of the diamond sample and laser sources. Both laser-illuminated confocal microscopy and brightfield microscopy operate independently of each other.

\section{OAM beam generation and characterization} \label{OAM beam generation}

\begin{figure*}[!h]
\centering\includegraphics[scale = 0.50]{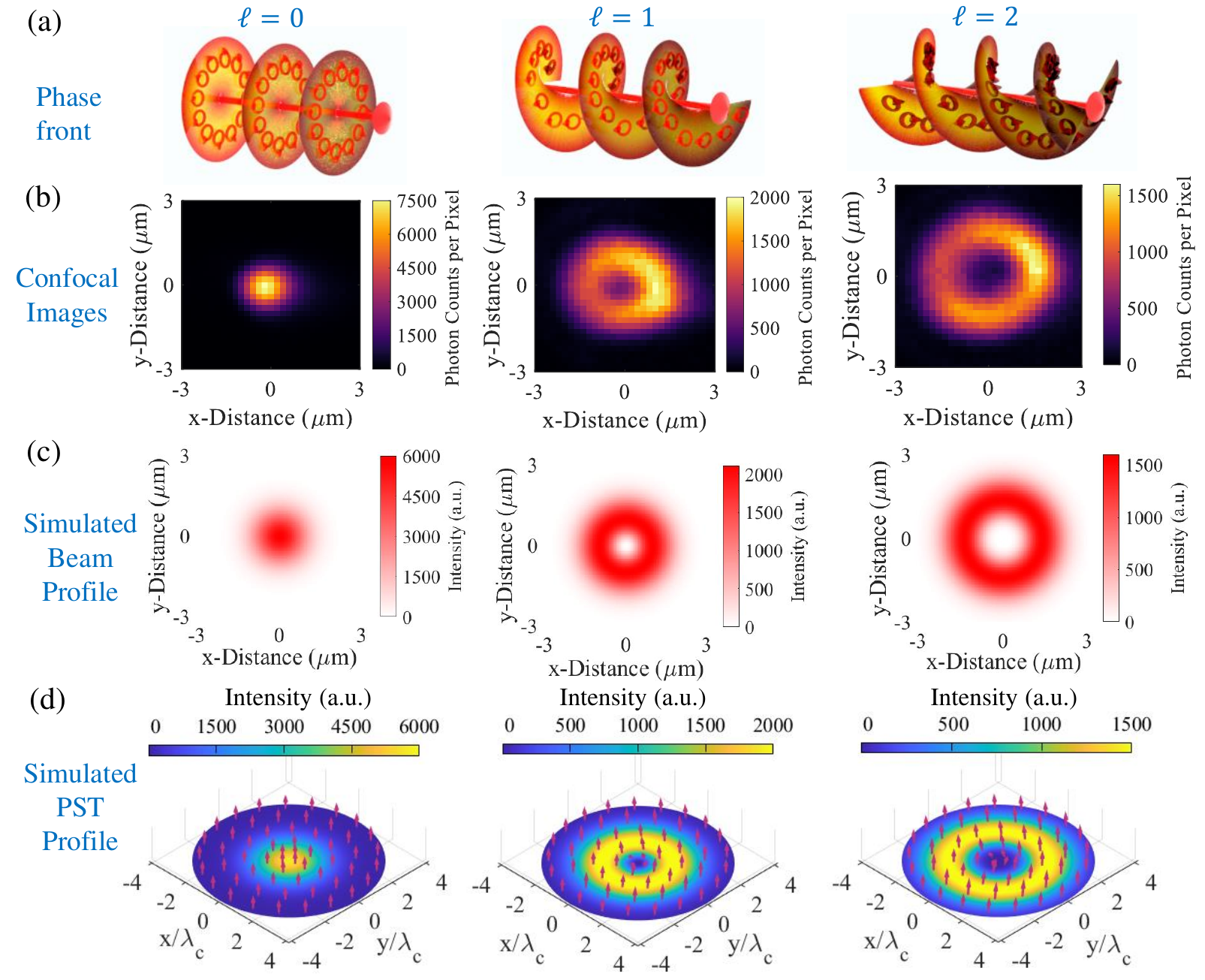}
\caption{Characterization and simulation of the intensity and spin texture profiles of the OAM beam. (a) The helical phase-front of the OAM beams with different topological numbers. (b) Experimentally obtained intensity-based confocal images of the OAM beam with different topological numbers, (c) Simulated intensity pattern using the parameters obtained from the confocal images, (d) Spin texture profile calculated from the beam intensity, parameters and the helical wavefront. }
\label{Figure_S2}
\end{figure*}

We perform a thorough characterization of the SLM-generated OAM beam prior to imaging with NV centers, to determine the beam properties and parameters. After the generation and alignment of the OAM beam, we acquire the intensity profile of the OAM beam with different topological numbers at the focal point of the objective lens using intensity-based confocal images (Fig. \ref{Figure_S2}(b,c)). These images are mainly based on the photon counts from the reflection of the OAM beams on the diamond surface. From these images, we can extract beam parameters and intensity profiles for different OAM numbers. Subsequently, using these parameters, we simulate the OAM beam to compare our experimentally obtained beam profile with the simulated one (Fig. \ref{Figure_S2}(c)). The electric field profile of an OAM beam is described by, 
\begin{equation}
{\cal E}(r,\alpha, 0) = {\cal E}_0\sqrt{\frac{2}{\pi \vert \ell \vert!}}\frac{1}{w_0}(\frac{r\sqrt{2}}{w_0})^{\vert \ell \vert} e^{-\frac{r^2}{\omega_0^2}}e^{i \ell \alpha}
\end{equation}
where ${\cal E}_0$ is the constant, $\ell$ is the OAM number, $w_0$ is beam waist diameter for $\ell=0$ ($\approx1.4\ \mu m$ for our case) and $\alpha$ is the spatial phase \cite{vallone2015properties}. The intensity profile is given by the term, $\vert{\cal E} (r,\alpha,0) \vert^2 $. We observe good agreement between the beam profile obtained from the confocal images and the simulated profile for different values of the OAM number, $\ell$ as shown in Fig. \ref{Figure_S2}. This confirms the correct focusing of the OAM beam on the diamond sample. Once, we get the the field profile of the OAM beam, we can use that information to simulate the spin texture in the beam. The PST is given by, $\vec{S_E}\propto\vec{\cal E}\times \vec{A}$. For monochromatic case, the electric field, $\vec{\cal E}$ can be written as, $\vec{\cal E} = Re(\vec{E}e^{-i\omega t})$. Now, taking $\vec{\cal E}=-\frac{\partial\vec{A}}{\partial t}$ and $\vec{E}=cos\theta\ \hat{x}+i\ sin\theta\ \hat{y}$, it can be shown that the magnitude of the PST is proportional to $\vert{\cal E}(r,\alpha, 0) \vert^2$. Considering the spin texture to be perpendicular to the plane of the wavefront, the orientation of the spin texture depends on the helicity of the OAM beam (Fig. \ref{Figure_S2}(a)). For a beam with OAM number $\ell$, the inclination of the wavefront with respect to the propagation axis of the beam is given by $\frac{\lambda}{2\pi r\ell}$, where r is the radial position about the beam axis. Fig. \ref{Figure_S2}(d) illustrates the spin texture of the OAM beam with $\ell = 0,\ 1$ and $2$ around the beam axis, where the strength, polarization, and direction of the spin texture are shown.

\section{Setup of the measurement protocols and parameters with NV centers} \label{Setup of the measurement protocols}

\begin{figure*}[!h]
\centering\includegraphics[scale = 0.42]{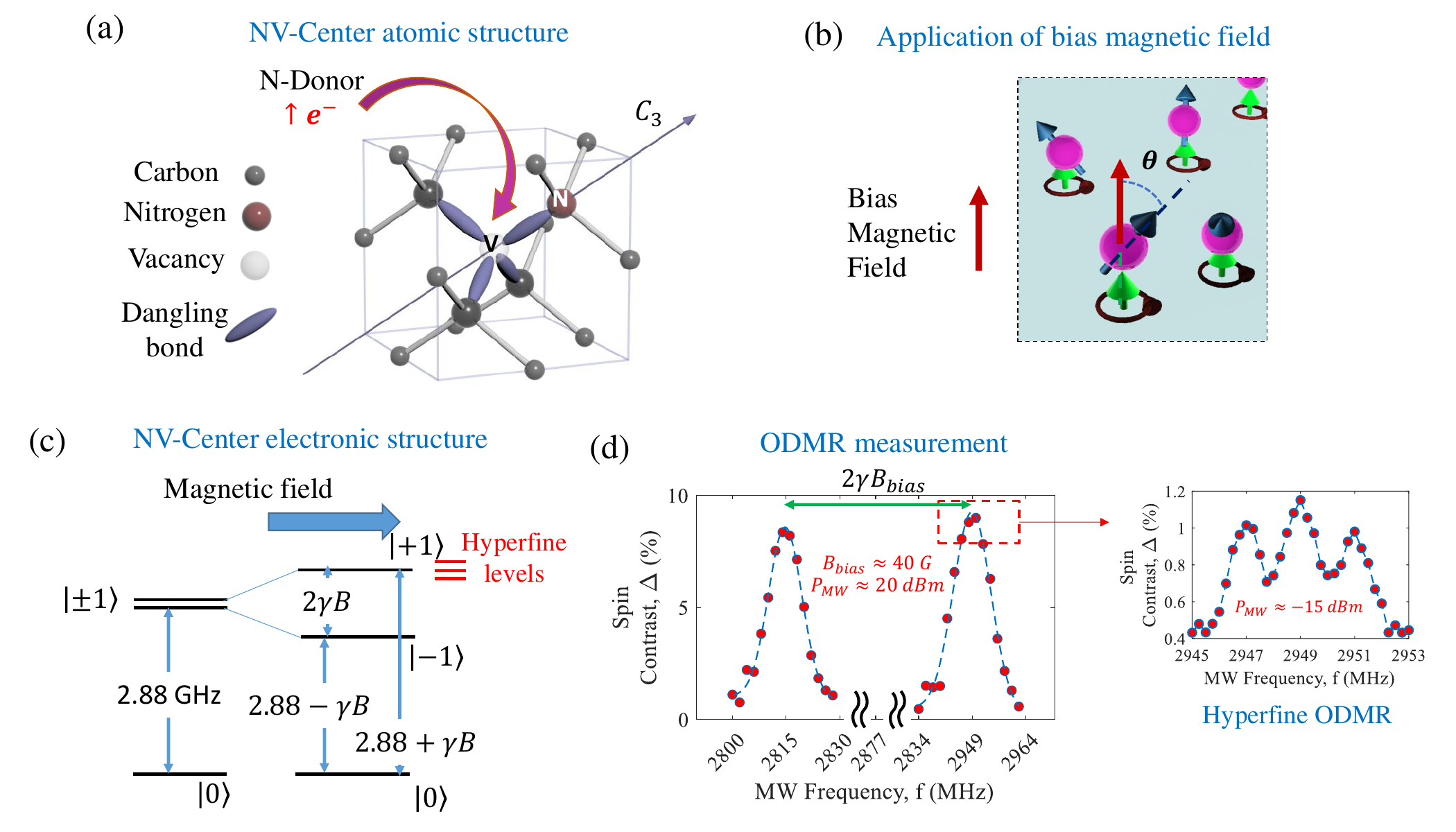}
\caption{Measurements for detecting the spin levels of NV centers in the diamond sample. (a) The atomic structure of NV centers showing their crystallographic orientation. (b) Biasing of the NV centers using an externally applied magnetic field. (c) Ground state spin energy levels of NV centers. Application of a magnetic field breaks the degeneracy of the $\vert\pm 1\rangle$ states. Hyperfine levels arise due to the interaction of the NV-center electron spin with the N-14 nuclear spin. (d) Optical detection of the ground state spin energy levels of NV centers.}
\label{Figure_S3}
\end{figure*}

We need to perform several characterization steps to build the experimental protocols for imaging the spin texture in an OAM beam in our setup. The first step is to accurately determine the spin energy levels of the ground states of NV centers. The diamond sample we are using is a bulk ensemble with an NV center concentration of around 300 ppb, oriented with a (100)-top surface. Due to the crystallographic orientation, the NV centers are oriented in four different directions (Fig. \ref{Figure_S3}(a)). The axes of all NV centers are at a $54.7^o$ angle with respect to the perpendicular direction of the diamond surface (Fig. \ref{Figure_S3}(b)). The ground state energy levels of NV centers consist of degenerate $\vert \pm 1\rangle$ states separated from $\vert0\rangle$ state by a zero-field splitting of around 2.88 GHz at 77 K. We apply a bias magnetic field perpendicular to the sample plane to lift the degeneracy of the spin levels, enabling each level to be addressed separately. This magnetic field affects all NV centers in different directions equally. Fig. \ref{Figure_S3}(c) illustrates the schematic of the electron spin energy levels of NV center ground states. The hyperfine splitting of each level arises from the interaction with N-14 nuclear spin. The positions of the spin energy levels can be accurately determined from the optically detected magnetic resonance (ODMR) measurements. Fig. \ref{Figure_S3}(d) presents the ODMR measurements at a bias magnetic field of around 40 Gauss. To precisely determine the position of spin levels, we can reduce the microwave power to observe the hyperfine peaks in the ODMR.

\begin{figure*}[!h]
\centering\includegraphics[scale = 0.5]{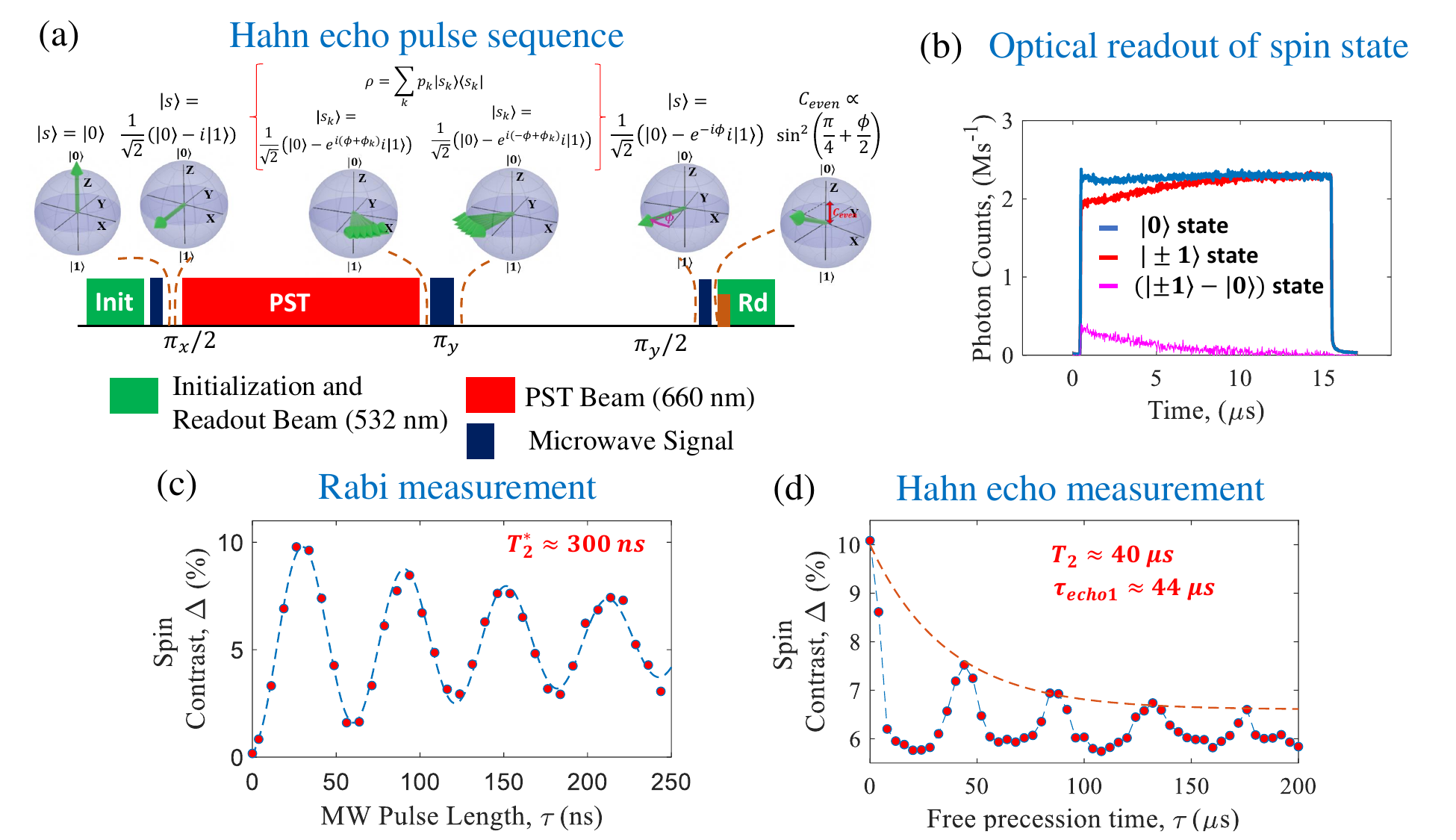}
\caption{NV spin characterizations to determine magnetometry measurement parameters. (a) Hahn-echo measurement protocol for measuring the PST-induced phase change of NV centers. (b) Measurement to determine contrast lifetime and readout pulse length. (c) Rabi measurement to find the lengths of $\pi-$ and $\frac{\pi}{2}-$pulses. (d) Hahn-echo measurement for finding the free precession time of the NV spin.}
\label{Figure_S4}
\end{figure*}

Once the spin energy levels are determined through the resonance measurements, we perform the AC magnetometry with the Hahn-echo pulse protocol to read out the PST-induced phase change of NV centers. Fig. \ref{Figure_S4}(a) depicts the pulse sequence and spin states for the Hahn-echo measurement protocol, where different parameters are determined through several characterization steps. The spin contrast lifetime and spin readout pulse length can be determined from the measurements in Fig. \ref{Figure_S4}(b). Here, the PL signal for $\vert0\rangle$ and $\vert1\rangle$ states are time-tagged with a resolution of $250\  ns$. The spin readout signal for any state is the photon counts difference with respect to that of $\vert 0\rangle$ state. From the graph, it is visible $\vert\pm1\rangle$ is completely initialized into $\vert 0\rangle$ state after $15\ \mu s$ of green laser excitation. Next, we perform the Rabi measurements to find the MW signal power and length needed to drive the NV center spin coherently between different states (Fig. \ref{Figure_S4}(c)). The MW pulses drive the NV electron sinusoidally from $\vert0\rangle$ to $\vert 1 \rangle$ states. From the measurements, we determine the length of $\pi-$pulse  based on the time corresponding to the max position of first oscillation. The length of $\frac{\pi}{2}$-pulse is approximately half of that of the $\pi-$pulse. Following this, we perform the Hahn-echo measurement protocol without applying the PST beam to determine the spin free precession time (Fig. \ref{Figure_S4}(d)). This measurement provides the position of the first revival peak and the $T_2$ spin coherence time of NV centers. For subsequent measurements of the NV center phase change due to the interaction with PST beam, the free precession time corresponds to the time of the first revival peak. The length of PST beam is adjusted to fit within this free precession time slot. This measurement pulse sequence is implemented to image the PST-induced phase change in NV centers at different positions of the OAM beam. Using this methodology, we can reach a NV phase sensitivity of around $40\  Degree/\sqrt(Hz)$. A summary of the different measurement parameters is provided in Table. \ref{Parameter_table}. 

\begin{table}[h!]
    \centering
    \begin{tabular}{c|c}
        Parameters &  Values\\
        \hline \hline
        NV free precession time  & $\approx$ 42 $\mu s$ \\
        PST beam length & $\approx$ 20 $\mu s$\\
        $T_2$  & $\approx$ 40 $\mu s$ \\
        $T_1$  & $\approx$ $5\ ms$ \\
        $\pi$$-$pulse length  & $\approx$ 30 $ns$ \\
        Maximum optical spin ellipticity  & $0.8$ \\
        Green laser Spot-Diameter  & $\approx 500\ nm$ \\
                Green laser Power  & $\approx 100\ \mu W$ \\
        \hline    
    \end{tabular}
    \caption{Parameters used for imaging the PST in an OAM beam by measuring the phase change of NV centers.}
    \label{Parameter_table}
\end{table}

\section*{References}
\bibliographystyle{iopart-num}
\bibliography{References}

\end{document}